\documentclass[a4, 12pt]{article}
    \usepackage[round]{natbib}
    \usepackage{longtable}
    \usepackage{geometry}
    \usepackage{array}
    \usepackage{multirow}
    \usepackage{wrapfig}
    \usepackage{pdflscape}
    \usepackage{tabu}
    \usepackage[normalem]{ulem}
    \usepackage{makecell}

    \usepackage{fullpage}
    \usepackage{amssymb}
    \usepackage{float}
    \usepackage{amsmath}
    \usepackage{bbm}
    \usepackage{mathtools}
    \usepackage{dsfont}
    \usepackage{enumerate}
    \usepackage{bm}
    \usepackage[Export]{adjustbox}
    
    \usepackage{booktabs}
    \usepackage{dcolumn} % Align on the decimal point of numbers in tabular columns
    \newcolumntype{d}[1]{D{.}{.}{#1}}
    \usepackage[flushleft]{threeparttable} % For better formatting of table notes
    \usepackage{threeparttablex}
    \usepackage{tabularx}
        \usepackage{cellspace}
    \usepackage[font=sc, labelsep=period]{caption}
    \usepackage{subcaption}

    \usepackage{graphicx}
    \graphicspath{{./graphs/}}

    \usepackage[pagebackref=true]{hyperref}
    \renewcommand*{\backref}[1]{}
    \renewcommand*{\backrefalt}[4]{%
        \ifcase #1 (Not cited.)%
        \or        [Cited on page~#2.]%
        \else      [Cited on pages~#2.]%
        \fi}
    \usepackage{url}
        
    \usepackage[utf8]{inputenc}

    \usepackage{scalerel,stackengine}
    \stackMath
    \newcommand\reallywidehat[1]{%
    \savestack{\tmpbox}{\stretchto{%
    \scaleto{%
        \scalerel*[\widthof{\ensuremath{#1}}]{\kern.1pt\mathchar"0362\kern.1pt}%
        {\rule{0ex}{\textheight}}%WIDTH-LIMITED CIRCUMFLEX
    }{\textheight}% 
    }{2.4ex}}%
    \stackon[-6.9pt]{#1}{\tmpbox}%
    }
    \usepackage{titlesec}
    \titleformat*{\section}{\normalsize\bfseries}
    \titleformat*{\subsection}{\normalsize\itshape}
    \titleformat*{\subsubsection}{\normalsize\itshape}
    \titleformat*{\paragraph}{\normalsize\itshape}
    \usepackage{wrapfig}  %to align figure on left or right
    \usepackage{verbatim}  %for skipping text be­tween \be­gin{com­ment} - \end{com­ment}
    \usepackage{lscape}  %to have horizontal pages

    \usepackage[table,xcdraw]{xcolor}

    \usepackage{tikz}
        \usetikzlibrary{calc}
        \usetikzlibrary{trees}
        \usetikzlibrary{decorations.pathreplacing,angles,quotes}
        \usepackage{pgfplotstable} 

    \usepackage{titling}
        \usepackage{tabularray}
        \UseTblrLibrary{booktabs}
        
        \NewTblrTheme{MyTheme}{
            \DefTblrTemplate{remark}{MyTmp}{
                \DefTblrTemplate{remark-tag}{MyTag}{}
                \DefTblrTemplate{remark-sep}{MyTag}{}
                \DefTblrTemplate{remark-text}{MyTag}{\InsertTblrRemarkText}
                \MapTblrRemarks{%
                \noindent
                \UseTblrTemplate{remark-tag}{MyTag}%
                \UseTblrTemplate{remark-sep}{MyTag}%
                \UseTblrTemplate{remark-text}{MyTag}
                \par
                }
            }
        
            \SetTblrTemplate{remark}{MyTmp}
        }

\usepackage{authblk}      % handles authors & affiliations
\usepackage[perpage,para]{footmisc}  % numeric footnotes instead of symbols

\title{Pull-Forward and Induced Vaccination Under Time-Limited Mandates: 
Evidence from a Low-Coercion Mandate}

\author[1]{Fabio I.\ Martinenghi%
  \thanks{Corresponding author: \texttt{fabio.martinenghi@newcastle.edu.au}. 
  We would like to thank Jason Abaluck, Basit Zafar, and Maryam Naghsh Nejad 
  for helpful comments and suggestions. 
  This project is funded by The Commonwealth of Australia - MRFF Scheme (MRF2019107). CCB is supported by a NHMRC Investigator Grant (APP1173163).
  MandEval is funded through Medical Research Future Fund [2019107]. 
  The authors thank Lorena Herrero, Research Manager, for her assistance. 
  We also acknowledge the broader contributions of the team members. 
  The authors thank Catherine Hughes for her support and guidance on MandEval. 
  HL is also supported by MRFF, project (Grant ID GA4147532024/GR000163); 
  and Western Australian's Future Health Research and Innovation Fund 
  (Grant ID WANMAIdeas2025-25/7) in collaboration with CB, BL, KA, and HM.
  }}

  \author[1]{Mesfin Genie}
  \author[3,4]{Katie Attwell}
  \author[4,5]{Huong Le}
  \author[4,5]{Hannah Moore}
  \author[1,2]{Aregawi G.\ Gebremariam}
  \author[9]{Bette Liu}
  \author[1]{Francesco Paolucci}
  \author[4,6,7,8]{Christopher C.\ Blyth}

\affil[1]{Newcastle Business School, University of Newcastle, Newcastle, NSW 2300, Australia}
\affil[2]{Department of Public Health, Policy and Systems, University of Liverpool, Liverpool L69 3GF, UK}
\affil[3]{VaxPol Lab, Political Science and International Relations, School of Social Sciences, The University of Western Australia, Perth, WA, Australia}
\affil[4]{Wesfarmers Centre of Vaccines and Infectious Diseases, The Kids Research Institute Australia, Perth, WA, Australia}
\affil[5]{Centre for Child Health Research, The University of Western Australia;
           and School of Population Health, Curtin University, Australia}
\affil[6]{School of Medicine, University of Western Australia, Perth, WA, Australia}
\affil[7]{Department of Infectious Diseases, Perth Children's Hospital, Perth, WA, Australia}
\affil[8]{Department of Microbiology, PathWest Laboratory Medicine WA, QEII Medical Centre, Perth, WA, Australia}
\affil[9]{School of Public Health and Community Medicine, The University of New South Wales, Sydney, NSW, Australia}

\date{}

\begin{document}
\maketitle

%-----------
\newpage
\begin{abstract}
    Vaccine mandates featuring a deadline, i.e. time-limited,
    can raise uptake either by
    pulling forward vaccinations that would have occurred later or by
    inducing additional vaccinations that would not have occurred absent
    the mandate. This paper asks how such mandates change vaccination behaviour,
    how the overall effect decomposes into the pull-forward and induction
    components, and which features of the mandate and public-health context
    drive that composition. Empirically, we study a low-coercion time-limited
    mandate targeting graduating high-school students in Western Australia and
    identify its causal effects using regression discontinuity designs based
    on strict school-age eligibility rules, applied to population-wide
    administrative records on first-dose COVID-19 vaccinations. We estimate
    both a static RDD at the deadline and a dynamic RDD that estimates the
    treatment effect over time. The mandate increased short-run first-dose
    uptake by 9.3 percentage points (12.7\%) among the targeted cohort,
    but the dynamic evidence shows that this effect is entirely driven by
    pull-forward behavior: uptake converges in the long run, implying
    no vaccinations were induced. Students advanced vaccination
    by up to 80 days. Theoretically, we develop a simple present-bias model
    of vaccination under deadlines. We use it to interpret the empirical
    patterns and to derive, among other results,
    conditions under which time-limited mandates are
    more likely to pull forward vaccinations rather than inducing them.
    Our findings highlight the importance of evaluating mandates beyond
    short-run windows and provide a framework for designing and interpreting
    time-limited vaccination policies.
\end{abstract}

\section{Introduction}

A vaccine mandate may aim to primarily pull forward
vaccinations that\textemdash absent the mandate\textemdash would have occurred
later, or induce new vaccinations that would otherwise
not have occurred \citep{brehm2022}, or both equally.
The former is also called ``intensive margin'' \citep{brehm2022}
or ``displacement'' effect\footnote{
    This terminology is mostly used to denote spatial displacement
    though, e.g., \cite{Buttenheim2022}},
and the latter ``extensive margin'' or ``net'' effect.
Because policy-makers may be trying to maximise one or the other effect
depending on the public health context, they would benefit from understanding
what are the mandate characteristics driving not only the overall uptake,
but also its pull-forward and induction components.
Pulling forward vaccinations leads to earlier immunity.
This is vital in the context of pathogens with strong seasonality
(e.g., influenza in winter) or fast transmission (e.g., COVID-19).
Conversely, inducing new vaccinations is most important, for instance,
in contexts where a high incidence of vaccine hesitancy is expected.

This paper asks \textit{how} time-limited vaccine mandates\footnote{
    These are vaccine mandates that include a deadline for vaccinating.
    In a strict sense, they include all policies that require vaccination within a
    certain date in order to access or participate in a one-off event.
    These events can be concerts, but also vaccine lotteries
    \citep{Campos-Mercade2024}. In a loose sense, they include any mandate that has
    an enforcement date\textemdash virtually any mandate. While our application
    belongs to the former group, our theoretical framework can also be applied
    to the latter.}
change vaccination behavior\textemdash
to what extent by inducing additional vaccinations and to what extent by
pulling forward vaccinations that would have occurred anyway\textemdash
and which features of the mandate and the broader public-health context
drive that composition.
This can have substantial policy relevance for two reasons.
First, understanding the causal links between
key variables such as the features of the mandate, the target population, and
the outbreak can help policy-makers design optimal mandates.
Second, optimality can only be defined based on a well-specified objective,
so that formally defining the policy objectives\textemdash not only uptake,
but its pull-forward and induction components too\textemdash
and modeling how they are shaped by the above variables can further aid
rigorous policy-design work.

We address this question both empirically, via a regression discontinuity
analysis of a vaccine mandate, and theoretically, by proposing a simple model of
(first-dose) vaccination under a time-limited mandate and subject to present
bias\textemdash exploring its implications both for our working example and
time-limited mandates in general.
Empirically, we study the effect of a low-coercion time-limited vaccine
mandate on vaccine uptake and decompose
this effect into the pull-forward and induction channels.
Because the mandate targeted only students graduating from high-school in
Western Australia (WA)\textemdash conditioning on vaccination
their access to a large-scale, national graduation celebration\textemdash and WA has
strict school-age laws, we are able to identify these effects via
regression discontinuity designs, which we apply to rich administrative data.
These WA-wide administrative records link date-of-birth,
location, and census data on school attendance to the universe
of first-dose COVID-19 vaccinations.
We take a standard (static) sharp regression discontinuity design (RDD)
approach and a dynamic RDD approach.
These allow us to capture, respectively,
short- and long-run impacts, thus disentangling the pull-forward
and the induction effects on vaccine uptake.

We find that the mandate increased the short-run vaccination rate by 9.3
percentage points (p.p.), or 12.7\%,
among students of the correct date-of-birth range to be in Year 12.
Our dynamic analysis shows that the short-run effect is entirely driven by
pulling forward future vaccinations, or equivalently, that the mandate did not
induce new vaccinations\textemdash nor discouraged vaccination.
Students responding to the mandate pulled
their vaccination forward as much as 80 days\footnote{
    This is true if one, as we do, takes population-level vaccination rates as
    the true rates. If one prefers taking a statistical inference approach,
    46 days is the correct value.
    Note that, for Year 11 and Year 12 students, vaccination rates were 90.5
    and 94.9 percent at 46 days since the policy deadline,
    and 98.2 and 98.7 percent 80 days since the policy deadline, respectively.},
making this policy an efficient tool for accelerating vaccination campaigns.

Theoretically, we develop a simple model of vaccination decisions under present
bias and time-limited mandates\textemdash an extension of
\cite{ODonoghueRabin1999}'s seminal work.
This model can be applied to any vaccine mandate that includes a deadline in
a strict sense\textemdash e.g., vaccination is required by a set date in
order to prevent termination from employment, or exclusion from one-off
events\textemdash but also in a loose sense, i.e. any mandate with an
enforcement date\textemdash enforcement dates act as deadlines.
In this latter sense, it covers virtually any vaccine mandate.
We first use our model to explain what likely generated the null-induction
result in our working example, then to explore, in general, how its variables
shape the size and composition of the uptake effect. We draw general lesson on
mandate design.

Using our model to interpret our empirical findings,
we suggest that the composition of the mandate's effect (a pure
pull-forward effect) is due to the epidemiological and policy
environment in WA at the time, namely, WA's policy
of re-opening its state borders only after reaching 90\% vaccine uptake
\citep{WA_Safe_Transition_2021},
which pushed long-run uptake upwards; and the low incidence of COVID-19 in WA,
which increased vaccine procrastination.
Using our model to study time-limited vaccine mandates in general, we find
that such mandates are more effective in the presence of:
(i) lower perceived risk of serious harm from not vaccinating;
(ii) conditional benefits of greater utility to the target
population and tied to the deadline;
(iii) greater present-bias, or greater
perceived immediate costs from vaccinating\footnote{
    These factors strengthen the mandate's impact only as long as they do not
    exceed the total vaccination incentives at the deadline (including the
    mandate's incentives).}; and
(iv) lower baseline incentives to vaccinate around the deadline.

The design of the mandate, the \textit{Leavers mandate},
is of independent interest for its features.
It ruled that WA graduating students had to get their first COVID-19 vaccine
dose to attend popular, annual post-graduation events in Dunsborough, WA,
which see about 9,000 students joining each year
\citep{schoolies2025, WestAust_VaxMandateAnnounce2021}.
This makes it a mandate with low coercion\textemdash as it excludes unvaccinated
individuals from a single event\textemdash and one that offers time-limited
non-monetary incentives.
In contrast, mandates such as
employment or university-level mandates \citep{Acton2024}\textemdash
which condition employment or university attendance on vaccinating\textemdash
can be considered as more coercive due to their longer-lasting
nature, and the magnitude of the economic consequences imposed on the
non-vaccinated.
Such mandates raise greater ethical issues and
can be costly for policymakers
from an economic and a political standpoint by,
respectively, reducing economic output and
generating political backlash and discontent \citep{bardosh2022}.

Our findings have three implications for policy.
First, mandates featuring low-coercion, non-monetary, and time-limited
incentives can be highly effective in accelerating vaccinations.
This aligns with predictions from behavioural economic theory, as also shown
via our model.
Moreover, under the right conditions (e.g., lower baseline incentive
to vaccinate), such mandates can induce new vaccinations too.
Second, it is important to evaluate a mandate's
effectiveness beyond the short-run. Failing to do so may result
in misinterpreting accelerated vaccinations as genuine increases in
long-term coverage.
Third, the epidemiological environment
(together with the beliefs people hold about it)
and the policy mix matter for the impact of this and other types of mandates.
Namely, we show how the expected benefits and costs from vaccinating,
and long-run incentives to
vaccinate, drive the impact of vaccine mandates on uptake
and their effect composition between
pulled-forward and newly-induced vaccinations.

We add to the literature using econometric methods to study
how vaccine mandates and other public-health interventions affect uptake.
Examples include \cite{Abrevaya2011}, showing that school-entry rules for
varicella increased uptake among children, \cite{Chang2016},
providing evidence that state insurance mandates increased infant
vaccination rates, and \cite{Carpenter2019} finding that mandating the take-up
of tetanus, diphtheria, and pertussis vaccine boosters among middle-schoolers
generated sizeable direct and spillover effects.
For hepatitis A, \cite{Lawler2017} finds that mandates outperform
non-binding recommendations.
These studies focus on the type of vaccine mandates that are generally referred
to as ``routine childhood vaccinations''.
Routine childhood vaccinations are preventive measures that are a
long-standing feature of public health systems worldwide and are broadly
accepted by the public. Instead, COVID-19 vaccine mandates were
emergency interventions implemented during a pandemic and
in response to a novel fast-spreading virus,
with very different social and political reception.

COVID-19 triggered a new wave of mandate research.
In their research protocol,
\cite{Gebremariam2025} lay out a plan to investigate the impact of Covid-19
vaccine mandates on vaccine uptake and
other outcomes using cross-country as well as individual-level Australian
administrative data.
The extant literature shows that
cash-lottery incentives raised first-dose uptake in Ohio \citep{brehm2022} and
across U.S. states \citep{BarberWest2022}; proof-of-vaccination (``green-pass'')
schemes increased uptake in Canada \citep{Fitzpatrick2023}; college mandates
curbed community spread \citep{Acton2024}; and sector-wide requirements affected
health-care utilisation and spending \citep{Aslim2024}.
Dynamic event-study work uncovers heterogeneous mandate impacts over time
\citep{Nguyen2024}, while micro-data from Indiana schools highlight large
indirect benefits from vaccinated peers \citep{Freedman2023}.

Our study is distinct from previous studies.
It is first in providing a formal framework for understanding the effect of
vaccine mandates and its channels. We hope this can be a useful reference to
both future researchers in this area and policy-makers.
In an epidemic context, it is first in
disentangling whether vaccine incentive programs lead
more unvaccinated people to be vaccinated or only pull forward vaccinations
through time, leading people who would get vaccines anyway to get them earlier.
While \cite{Campos-Mercade2024} do disentangle the two effects,
they study an intervention that targets potential booster-dose
recipients, and hence excludes unvaccinated individuals from consideration.
Instead, we focus on a population of unvaccinated individuals, and thus
address the critical public health question of whether a policy can
make the unvaccinated vaccinate.
We do this thanks to a long panel of administrative data
coupled with a dynamic RDD approach.

Empirically,
even in the broader literature about the impact of vaccine policies on the
uptake of any vaccine, studies disentangling the above effects remain
scant, with \cite{Carpenter2019} and \cite{Abrevaya2011} being notable
exceptions.
Other contributions include providing the first application of dynamic RDD to a
health setting, and the first impact estimates of a mandate that uses
time-limited non-monetary incentives.

The remainder of the paper proceeds as follows. Section \ref{sec:background}
describes the institutional setting; Section \ref{sec:data} the data;
Section \ref{sec:methods} outlines the baseline and dynamic RDD specifications;
Section \ref{sec:results} presents the main findings and robustness checks.
Section \ref{sec:ev} presents our simple theoretical model, derived in full in
Appendix \ref{app:ev}.
Finally, Section \ref{sec:conclusion} provides policy implications and
concludes.

\section{Background\label{sec:background}}

In late 2021, Western Australia (WA) recorded virtually no local COVID-19
transmission.  As of 6 October, the state had just 15 active cases out of 1,110
confirmed infections and nine deaths since the pandemic began,
with 1,788,405 tests conducted to date \citep{HealthyWA_Case_6Oct2021}.
WA's stringent border closures and internal restrictions insulated the
population from community spread.  National Cabinet's ``National Plan'' linked
reopening to vaccination thresholds, and the WA Government announced it would
not ease its hard border until at least 90 percent of eligible residents
were fully vaccinated
\citep{Guardian_ReopeningRoadmap2021, WA_Safe_Transition_2021}.
WA's successful elimination strategy left some West Australians perceiving
minimal personal risk from COVID-19 and/or feeling like they wanted to wait
longer, despite the target for reopening \citep{Carlson2022}.
This dynamic depressed the demand for vaccination\textemdash
including among adolescents and their parents\textemdash
alongside concerns about the newness of the
vaccine and worries about side effects \citep{Carlson2023}.
It is noteworthy that adolescents in WA can consent to their own
vaccinations (including COVID-19 vaccinations)
if they are aged 16 or older, covering both our treatment
and control group, and otherwise need parental consent.

The state's COVID-19 Vaccination Program rolled out in phases.
Phase 1a of the overall program began on 22 February 2021 with frontline workers.
Subsequent phases expanded eligibility to older adults,
people with comorbidities, and then to all individuals aged 12 and over,
with the pace managed by a joint Commonwealth-State
implementation plan
\citep{WA_VaccineProgramPhases2021,WA_VaccinationProgramAudit2021}.
By 21 October, WA had administered at least one dose to 85.8 percent of
those aged 16 and over \citep{DH_AustraliaRollout20211021}.

\subsection{The Leavers Event \& its vaccine mandate}

The annual Leavers celebration in Dunsborough is Western Australia's largest
sanctioned event for graduating Year 12 students.
Official Leavers festivities run each year in late November
and were expected to draw about 10,000 graduates from across the state
in 2021
\citep{WA_LeaversProof2021}.  Attendees converge on Dunsborough\textemdash
a coastal town in the South West\textemdash
to mark the end of secondary school with concerts, beach parties,
and community events \citep{TribeTravel_LeaversInfo2021}.

On 1 October 2021 Premier Mark McGowan signalled that requiring a
first vaccine dose for Year 12 students to attend Leavers events was a
``strong possibility” under consideration
\citep{McGowan_Oct1_2021,turn0search0,Guardian_ReopeningRoadmap2021}.
By 5 October, WA Vaccine Commander Chris Dawson
confirmed that ``before you get your wristband you will have to provide
evidence of vaccination'' to gain entry to official school leavers celebrations
\citep{Carmody_Oct5_2021,turn1search0,ABC_DunsboroughLeavers2021}.
The formal policy announcement came on 15 November 2021: all participants
in the 22-25 November Dunsborough Leavers event were required to show proof
of at least one COVID-19 dose via the Leavers WA app
by 21 November 2021
\citep{WestAust_VaxMandateAnnounce2021, WA_LeaversProof2021}.

Finally, while being unvaccinated barred you from entry to the Leavers
celebrations, unvaccinated individuals found on the Leavers grounds
were punishable by a fine of up to AUD 20,000 for individuals and
AUD 100,000 for staff members.

\subsection{School age law in WA}
In Western Australia, compulsory schooling begins with Pre-primary, and
a child is eligible to start Pre-primary in a given year only if they will have
turned five on or before 30 June of that year.
Children who reach five after that date begin the following
year \citep{WAEnrollingSchool}.
As a result, children typically start pre-primary between ages
four-and-a-half and five-and-a-half, depending on their date of birth,
Year 1 between five-and-a-half and six-and-a-half, etc.
Following pre-primary, students continue through twelve additional years of
schooling, with high-school years ranging between Year 7 and Year 12,
the latter being approximately thirteen years after their initial
enrollment in pre-primary \citep{WAEnrolmentPolicy}.

The cutoff date of 30 June is a strict state policy and
enforced tightly by principals \citep{ABC2019}.
A parent wanting to defer their child's
entry may be asked to present the principal with expert evidence\textemdash
such as reports from a paediatrician, psychologist or other
specialist\textemdash demonstrating that the child's developmental needs
(for example significant medical or developmental delays, or the aftermath of a
traumatic event) would make standard enrolment detrimental to the child's
development.
This makes the WA cohort different from that of some other states, such as
New South Wales, where parents can defer the enrolment of their children by up
to one year \citep{nsw_education_act_1990}.

The strictness of this policy is also reflected in
Figure \ref{fig:deadln}, which shows
a clear separation between all the vaccination rates of students in the
age-range for Year 12,
the target population, and those in the age range for Year 11.

\section{Data \label{sec:data}}
We conduct this study using the Australian Immunisation Register linked to the
Person Level Integrated Data Asset (AIR-PLIDA).
The AIR provides comprehensive vaccination records, and we use it to identify
first COVID-19 vaccine doses, their timing, and their recipient.
We link AIR data to a selection of PLIDA products: (i)
the Core Location data asset from the Australian Bureau of Statistics (ABS),
to exclude individuals who were not residents of Western Australia at the time
of the vaccine mandate; (ii) the Core Demographics data asset to obtain precise
month-year birth dates for all WA residents;
(iii) we use the 2021 Census to identify students.
The resulting dataset is at the individual level.

We rely on answers to Census about school attendance (variable \textit{TYPP}).
In particular, in our main analysis, which focuses on students, we
exclude from our sample those individuals who replied to the question
``What type of education institution is Person \textit{X} attending?'' with
anything other than ``Secondary'' (and its subtypes).
Then, we use month-year of birth to identify which school year
enrolled students were attending in 2021.
This is an effective approach thanks to the aforementioned strict
policies on school starting ages
in WA, which in turn implies that date of birth (almost) deterministically
assigns to the treatment and control groups in our sample.
In sum, our inclusion criteria are (i) WA residents, (ii) students, and
(iii) born between July 1st 2003 and June 30th 2004, the treated group, or
July 1st 2004 and June 30th 2005, the control group.
Note that, because the threshold for starting school is the last day of
June, then the granularity of our date of birth data (month-year level)
does not introduce measurement error.

\begin{figure}[!hp]
    \includegraphics[width = \textwidth]{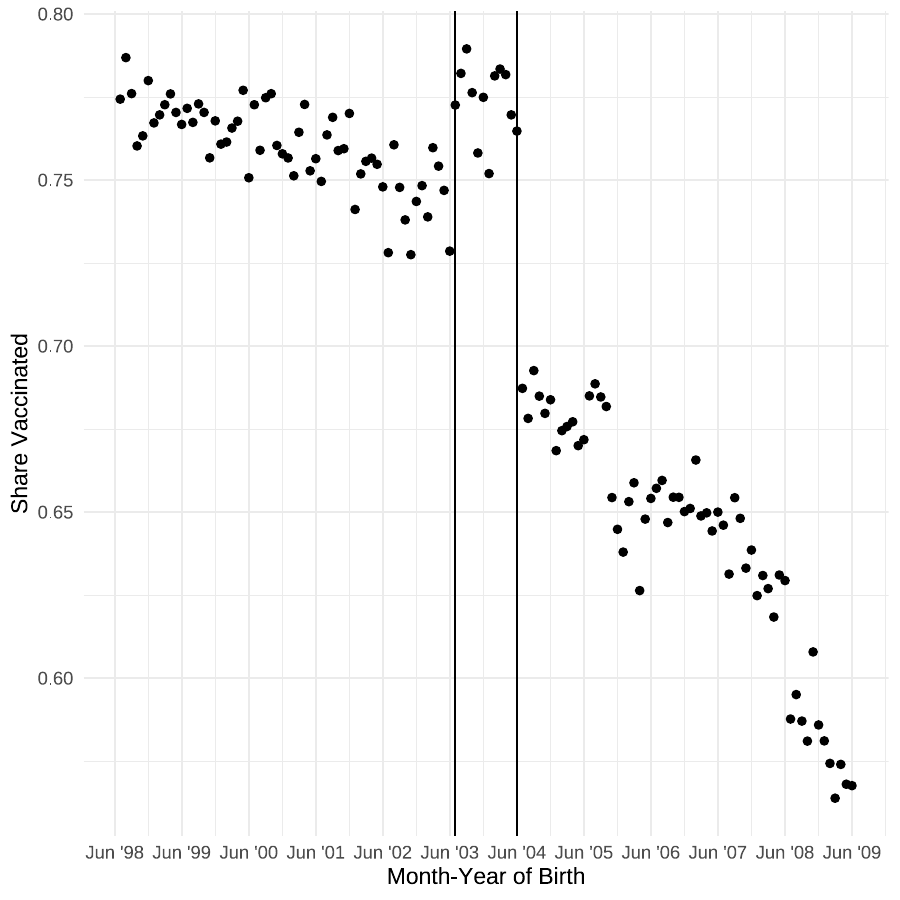}
    \caption{Vaccination rates in WA at the policy deadline by date of birth (at the month-year level)}
    \label{fig:glob}
    {\footnotesize
        \textit{Notes.} This figure plots
        the population-wide vaccination rate by month-year cohort and
        at the time of the mandate's deadline for vaccinating, for
        individuals born between June 1998 and June 2009.
        Equivalently, the denominator counts the number of
        WA residents within a given month-year cohort, while the numerator
        counts those who, among them, vaccinated by the policy deadline,
        21 November 2021.
    }
\end{figure}

Indeed, the treatment assignment rule is so sharp that
the effect of the \textit{Leavers mandate} on vaccine uptake
can be clearly spotted even by looking at the general population, without
focusing on students. We show this in Figure \ref{fig:glob}, where we
plot the unconditional vaccination rate (the denominator is the number of
WA residents within a given month-year cohort) by month-year cohort and
at the time of the mandate's deadline for vaccinating.
This shows a sharp jump in vaccination rates between people born from
July 2003 to June 2004\textemdash born in the right date range to be graduating
high-school in 2021\textemdash and those born outside this date range\footnote{
    We do not use the June 2003 discontinuity as individuals to the left of that
    cutoff are much more heterogeneous, including tertiary-level students and
    workers. Workers, in particular, were targeted by other (employment) vaccine
    mandates around the same time of the Leavers mandate announcement.
    This would confound treatment effects estimated at that cutoff.
}.

Our outcome is a binary indicator equal to one if the individual has received
the first dose of a COVID-19 vaccine, and zero otherwise.
In our primary analysis, we restrict the sample to the student
population, and we extend it to all WA residents as a robustness check in
Tables 1 and 2 in the Appendix. By construction, treatment effect estimates
from this latter sample are attenuated, but can be useful to place the policy
effects in the context of the whole WA population. Finally, all our
specifications focus on WA residents, as identified by the Australian
Bureau of Statistics using multiple government databases
(Core Demographics data asset).

\section{Empirical Strategy \label{sec:methods}}

RDD approaches are arguably the only sensible choice in this context.
They require minimal assumptions to identify treatment effects and these have
testable implications.
They exploit the design of the mandate, which targets a population
defined by their date of birth and their high-school graduation status.

Specifically, sharp RDD relies on the continuity of the conditional mean
functions of potential outcomes at the cutoff \citep{hahn2001}.
When the running variable is recorded in coarse intervals or has only a few
distinct values near the threshold,
it becomes difficult to verify or plausibly maintain this condition.
In such cases, a local-randomisation approach is often used,
where observations falling within a narrow, symmetric window around the
cutoff are treated as if randomly assigned to treatment or control,
turning the RD into a finite-sample experiment and where
inference is conducted via permutation (i.e. Fisherian) randomisation
instead of local-polynomial methods \citep[for a review of regression
    discontinuity designs, see][]{cattaneoReview}.

In our setting, either a continuity-based or a local-randomisation approach to
RDD could be justified.
Our running variable, date of birth, is encoded at the month-year
level, so that there is some granularity in the data\textemdash pointing to
a local-randomization approach\textemdash but not enough to make a
continuity-based approach questionable.

While this choice is not particularly consequential\textemdash as can be
seen by comparing Tables \ref{tab:rdd_cont} and \ref{tab:rdd_randinf}
in the Appendix\textemdash we prefer the continuity approach for
two reasons. First, to ensure that we are not giving excessive
weight to the observations in principle most vulnerable to manipulation
(here in the form of parents managing to delay or push forward the entry
to school of a child, which we cannot entirely rule out).
This is an \textit{a priori} argument,
as the \cite{mccrary08} test provides evidence
against the presence of a discontinuity in the density function of the
running variable around the cutoff.

For the same reason, we also make use of the full one-year window around the
cutoff, set at 30 June 2004, and give each observation equal weight
(uniform kernel).
In other words, we are comparing the vaccination rates of the
full 2021 Year 12 class (treatment group) with the 2021 Year 11
class (control group)\textemdash
those born July 2003-June 2004 with those born July 2004-June 2005
who all have a census record of being a student.

Second, a local-randomisation approach
compares outcome conditional means at two neighbouring intervals of the
running variable. Intuitively, it draws lines with no slope\textemdash
horizontal. Instead, the continuity-based approach estimates a local polynomial,
which can, hence, follow the slope of the density of the outcome conditional on
the running variable.
This flexibility is desirable here because vaccination uptake shows a clear
cohort trend (see Figures \ref{fig:glob} and \ref{fig:deadln})
and the polynomial can capture that slope while still identifying
the jump at the threshold.

Finally, we present below a formal description of our sharp RDD application.
Let $X_i$ denote student $i$'s month--year of birth and let the cutoff be
$c=\text{30 June 2004}$. Define the treatment-assignment indicator
$T_i \equiv \mathbbm{1}\{X_i \le c\}$,
i.e. equal to one if $i$ is in Year 12 in 2021 and zero otherwise.
Let $Y_{i,t}$ be an indicator that $i$ has received (at least) the first
COVID-19 dose by calendar date $t$.
Our estimand is the sharp RD effect at date $t$,
\[
    \tau_{\text{SRD}}(t)
    \;\equiv\;
    \lim_{x \uparrow c}\,\mathbb{E}\!\left[\,Y_{i,t}\mid X_i=x\,\right]
    \;-\;
    \lim_{x \downarrow c}\,\mathbb{E}\!\left[\,Y_{i,t}\mid X_i=x\,\right],
\]
which we estimate via local linear regression with a uniform kernel on a fixed
12-months window, $[c-12,\,c+12]$. Identification of $\tau_{\text{SRD}}(t)$
follows \cite{hahn2001}.

The assumptions required for identification are:
(i) right- and left-hand limits of the relevant conditional
expectations exist in a neighbourhood of $c$;
(ii) the density of $X_i$ is positive at
$c$ (no precise manipulation, which we check via a McCrary test);
and (iii) the conditional mean functions $\mathbb{E}[Y_{i,t}(0)\mid X_i=x]$ and
$\mathbb{E}[Y_{i,t}(1)\mid X_i=x]$ are continuous at $c$, where
$Y_{i,t}(0)$ and $Y_{i,t}(1)$ denote the potential vaccination outcomes at
time $t$ if student $i$ were assigned to the Year 11 and Year 12 sides of the
cutoff, respectively.

A minor but relevant point is that the Leavers mandate actually targeted,
and hence \textit{treated}, only those 2021 Year 12
students who both graduated and intended to attend the official events, rather than
the whole 2021 Year 12 cohort.
Let $D_i$ be the indicator identifying the students actually treated
by the mandate. Then, the local average treatment effect at $c$
of the Leavers mandate for the complier population\footnote{
    In our fuzzy RD, the running variable is date of birth and the
    instrument $T_i$ is an indicator for being on the Year--12 side of the cutoff.
    The treatment $D_i$ indicates belonging to the group targeted by the Leavers
    mandate, that is, successfully completing Year 12 in 2021 and planning to
    attend the official Leavers events in Dunsborough. Compliers are
    defined by their treatment response to the instrument:
    $D_i(1)=1$ and $D_i(0)=0$. Intuitively, these are students who would be
    Leavers-eligible and planning to attend if they are assigned to Year 12 by the
    cutoff, but would not be in the mandate-targeted group if they are instead in
    Year 11. Our fuzzy RD estimand can therefore be interpreted as the average
    effect of Leavers-mandate eligibility on vaccination by the deadline among
    these compliers.
} is identified by the fuzzy RD estimand
\[
    \tau_{\text{FRD}}(t)
    \;=\;
    \frac{
        \tau_{\text{SRD}}(t)
    }{
        \lim_{x \uparrow c}\mathbb{E}[D_i\mid X_i=x]
        \;-\;
        \lim_{x \downarrow c}\mathbb{E}[D_i\mid X_i=x]
    }\,,
\]
under the usual assumptions of nonzero first stage, continuity of potential
outcomes at $c$, no precise manipulation of the running variable at $c$, and
monotonicity.
Because $D_i$ is unobserved in our data, we report $\tau_{\text{SRD}}(t)$ as the policy's
intention-to-treat (ITT) effect. We note that, since the denominator in $\tau_{\text{FRD}}(t)$\textemdash the
\textit{first stage}\textemdash is a jump in probability, then ITT is attenuated toward zero. In
particular, if the treatment effect at the cutoff is nonnegative, then
$\tau_{\text{SRD}}(t) \le \tau_{\text{FRD}}(t)$, i.e.,
the ITT underestimates the fuzzy-RD effect.

We estimate the impact of the Leavers mandate both in the short and long run.
For the short run, we estimate a sharp RD in assignment using date of birth as the running
variable and 30 June 2004 as the cutoff, evaluating vaccination status at the mandate
deadline, 21 November 2021.

For the long run, we re-estimate this RD on each calendar day, starting on
1 June 2021\textemdash four months before the policy was first
mentioned\textemdash and ending on 21 May 2022,
six months after the mandate deadline.
We keep the same date-of-birth cutoff and $\pm 12$ months (24-month total)
window in each RD instance, and compute standard errors via clustered multiplier
bootstrap.

\section{Results \label{sec:results}}

\begin{figure}[h]
    {\centering
        \subfloat[At policy deadline]{\label{fig:deadln}\includegraphics[width=.75\linewidth]{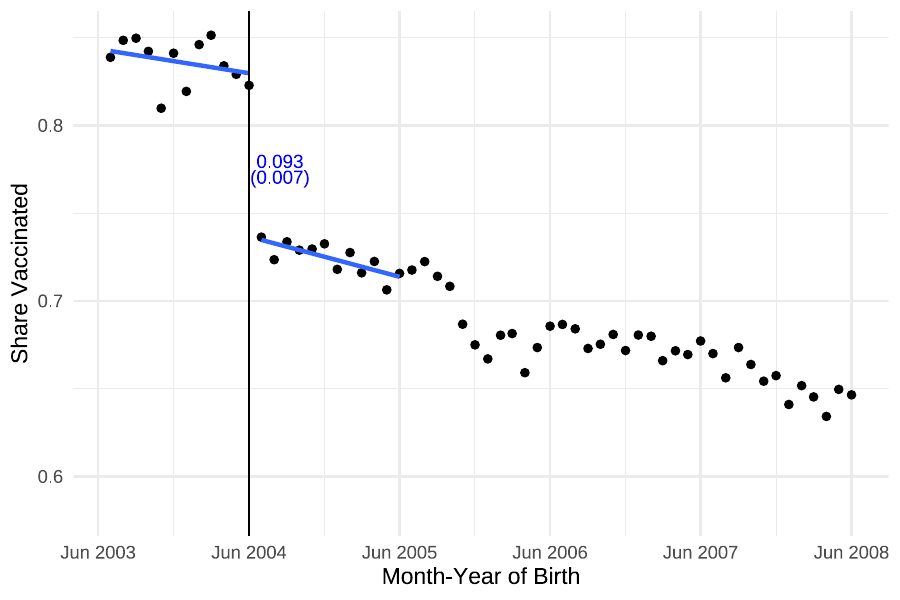}}\hfill \\
        \subfloat[At first policy mention]{\label{fig:frst_mention}\includegraphics[width=.75\linewidth]{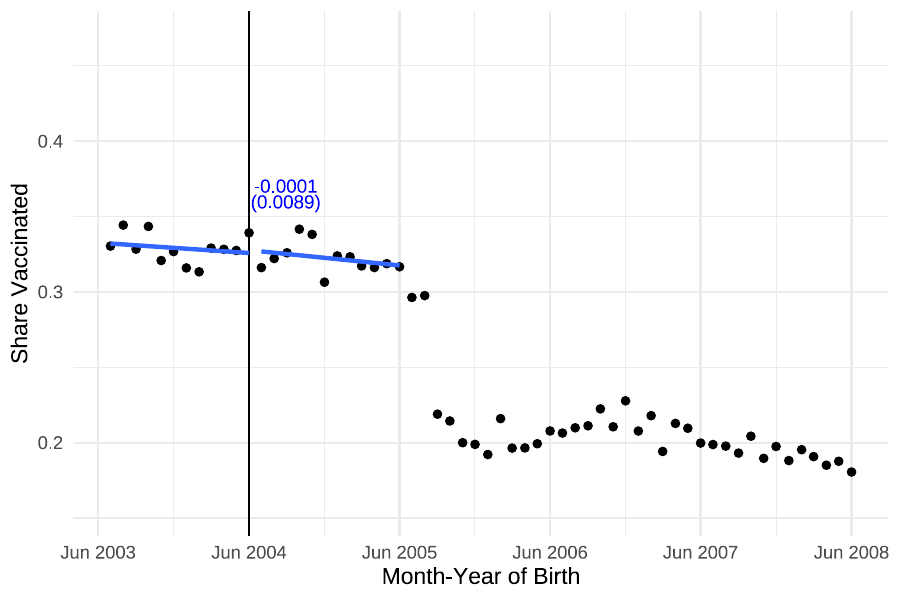}}\par
    }
    \caption{Student vaccination rates in WA}
    \smallskip
    {\footnotesize
        \textit{Notes.} This figure plots
        the vaccination rate of WA students by month-year cohort (a)
        at the time of the mandate's deadline for vaccinating
        (21 November 2021), and
        (b) at the date of the first public mention of the policy
        (1 October 2021). Students are identified via 2021 Census.
        Equivalently, the denominator counts the number of
        WA students within a given month-year cohort, while the numerator
        counts those who, among them, vaccinated by the given date.
    }
    \label{fig:sharp_rdd}
\end{figure}

\begin{figure}[!hp]
    \includegraphics[width = \textwidth]{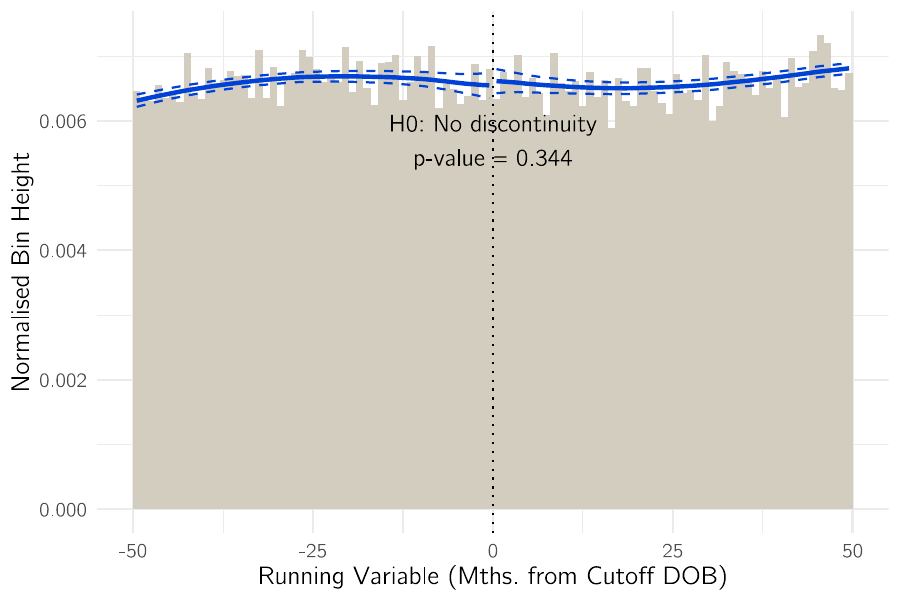}
    \caption{McCrary test for continuity in the density of the running variable at the cutoff}
    \smallskip
    {\footnotesize
        \textit{Notes.} This figure plots
        the density of the running variable, date of birth (DOB)
        at the month-year level, in grey. Following McCrary (2008), we
        separately fit one local linear regression for each side of the cutoff,
        June 2004. The p-value tests the hypothesis of
        ``no discontinuity at the cutoff''.
    }
    \label{fig:mccrary}
\end{figure}

\begin{figure}[!hp]
    \includegraphics[width = \textwidth]{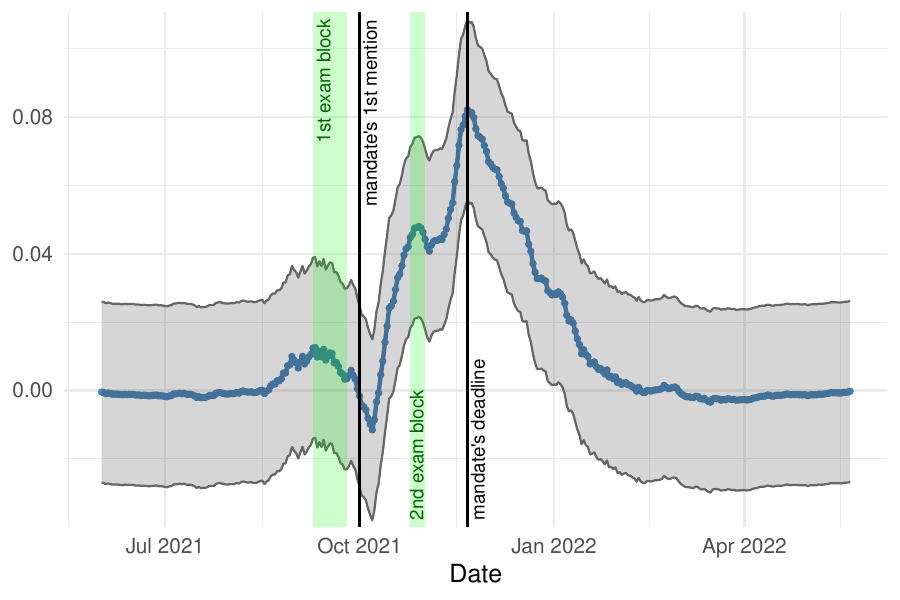}
    \caption{Dynamic RDD treatment effects of the policy}
    \smallskip
    {\footnotesize
        \textit{Notes.} This figure plots dynamic RDD estimates for the effect
        of the \textit{Leavers} mandate on vaccination rates.
        We run one sharp RDD for each date displayed and plot the
        treatment-effect estimates (dark blue) along with their
        bootstrapped 95\% confidence intervals. As for the baseline analysis,
        we set the bandwidth to 12 months of birth from the June 2004 cutoff
        and use a uniform kernel,
        hence comparing students that, in 2021, were attending
        Year 12 (treated) and Year 11 (control).
        We highlight in green the time intervals when final Year 12 exams were
        held, and draw black vertical lines for the key dates.
    }
    \label{fig:dyn_rdd}
\end{figure}

We find that the \textit{Leavers mandate} had a short-run effect of 9.3 p.p.
(s.e. 0.008) on the vaccination rate of Year 12 students, as shown in
Figure \ref{fig:deadln}.
This effect is the effect of the policy at vaccination deadline set by the
mandate as a condition to access the party.
As mentioned in Section \ref{sec:methods},
this likely underestimates the causal effect of being targeted by the Leavers
mandate.
We offer three pieces of evidence to support the validity of our findings.
First, the McCrary test (Figure \ref{fig:mccrary}) cannot reject the null
hypothesis of ``no discontinuity in the density of the running variable (here,
date of birth) at the cutoff''. In other words, we should not be worried
about one-sided manipulations of the date of birth of students, which would
invalidate our design.
Second, the null RDD estimate (coef. -0.01 p.p., s.e. 0.009)
at the time of the first mention of the mandate by a
government official, 1 October 2021, serves as supporting evidence that the
policy is responsible for the estimated effect
(see Figure \ref{fig:frst_mention})\footnote{
    Another discontinuity can be observed in Figure \ref{fig:frst_mention},
    approximately between
    Year 11 students and younger students. This was generated by the vaccination
    guidelines in place,
    which had not recommended vaccinating individuals 15 years old
    and younger until shortly before 1 October 2021. At that time,
    this cohort was just starting to vaccinate and, as can be seen in
    Figure \ref{fig:deadln}, eventually caught up.
}.
Third, we extend our analysis dynamically, running the same RDD every day between
1 June 2021 and 21 May 2022, and bootstrapping our standard errors.
The resulting estimates are plotted in Figure \ref{fig:dyn_rdd}.
This is further evidence that the estimated effect should be attributed to the
mandate as the ``pre-trends''\textemdash the estimated treatment effects before
the first mention of the mandate\textemdash are zero.

In the dynamic analysis, we also find that the effect converges to zero in
about one month since its peak on the deadline date. This means that the
short-run effect is entirely due to pulling forward future vaccination:
absent the policy, the students driving this effect would have still
vaccinated, but would have done so later. Indeed, we estimate that students
responding to the mandate pulled the vaccination forward between 46 and 80 days,
depending on whether one prefers taking a statistical inference approach or
takes population-level vaccination rates as the true rates.
Intuitively, we reach this conclusion after observing that Year 11 (control)
and Year 12 (treated) students had similar vaccination rates before the
policy announcement and after the treatment effect tapered off, indicating
that while the policy boosted vaccination among Year 12 students,
eventually Year 11 students caught up.

Moreover, inspecting the dynamic estimates, one may be concerned that the
conclusion of the exams themselves, not the mandate, are behind the surge in
vaccinations.
The logic of the concern is that a Year 12 student planning to sit
the exams might have decided, or their parents might have advised,
to delay the vaccination until after the exams. In this scenario, the
risk of experiencing a minor side effect that compromised exam performance
would have been seen as more likely than the risk of catching COVID and
experiencing this form of disruption instead. Such a risk calculation
would be consistent with the fact that there was no community transmission
of COVID in WA at the time and had not been for most of the prior two years
\citep[e.g., see][]{wahealth2021sep25, wahealth2021nov01}.
We show below that such concern is unlikely to be a significant driver
and very unlikely to be the key driver, and present supporting evidence.

The timing of the exams is marked in green in Figure \ref{fig:dyn_rdd},
showing that one phase occurs before the mandate is first mentioned, while
another occurs between first mention and mandate deadline.
Upon inspection of Figure \ref{fig:dyn_rdd}, we can see that the first
exam phase, involving ATAR course written examinations
\citep{scsa2021} and scheduled between 1 and 19 November 2021,
seems to cause a minor and short-lived adjustment in vaccination rates.
Its local maximum occurs on the day after the start of this phase, it is small
(1.26 p.p.), not statistically significant at the 95\% level, and it
is followed by a similarly-sized trough (-1.15 p.p.), also not statistically
significant at the 95\% level.
The other phase, involving ATAR course practical examinations
\citep{scsa2021}, was scheduled from 25 September to 17 October 2021.
A similar pattern ensues, with a small inflection in the context of a
strong trend, which continues and peaks on the deadline set by the mandate.
These findings should reassure that the conclusion of the exams
themselves is not a significant driver of vaccine uptake.

\section{A model of vaccination decisions \label{sec:ev}}

In this section, we develop a simple model of vaccination under present-bias
that explores the relationship between the
characteristics of the mandate design, its target population, and
epidemiological environment, on the one hand, and the effectiveness of the
mandate\textemdash both as a whole and in its induction and pull-forward
components\textemdash on the other hand.
We extend \citet{ODonoghueRabin1999} to $T$ periods and apply it to
model individual behaviour when faced with the decisions of whether and when
to take the first vaccine dose under a time-limited vaccine mandate.
The time-limited mandate adds a one-off payoff $M$
(for example, access to a memorable event) at a deadline $T_0$,
and individuals $i$ have quasi-hyperbolic preferences with
present-bias $\beta_i\in(0,1]$ and discount factor $\delta=1$
\citep[all results extend to $\delta<1$; see][]{ODonoghueRabin1999}.

When making the decision to take their first vaccine dose, individuals take into
account: $\lambda_s$, the period-$s$ health hazard\textemdash
severe illness or death\textemdash faced by the unvaccinated;
the effectiveness of the vaccine in reducing this hazard, $e\in[0,1]$;
and the harm associated with each adverse event, $H>0$.
They also take into account the non-mandate, non-health per-period benefits
$b_s$ that accrue to being vaccinated by calendar time $s$
(e.g., travel convenience); mandate's deadline $T_0\le T$,
and the mandate's one-off payoff $M>0$, which is conditional on the
vaccination occurring by period $T_0$.

At each decision time $t$, individuals compare vaccinating now with
waiting one more period.
Formally, Lemma~1 in Appendix~\ref{app:mandate} shows that they vaccinate at $t$
if and only if
\begin{equation}
    eH\lambda_t + b_t + M\,1_{\{t = T_0\}} \;\ge\; Z_i \;\equiv\; \frac{1-\beta_i}{\beta_i}\,c_i .
    \label{eq:decision-rule-main-text}
\end{equation}
The left-hand side collects the one-period incremental advantage of vaccinating
now rather than waiting, while the right-hand side captures the
individual-specific threshold, $Z_i$, implied by present bias $\beta_i\in(0,1]$
and the immediate cost $c_i$.
For brevity, we omit the $i$ subscript and work with the distribution of $Z$
in the population, denoted $F_Z$.
This heterogeneity in costs and present bias across individuals, modelled
as distribution $F_Z$, leads to heterogeneous baseline vaccination
decisions and heterogeneous responses to the vaccine mandate.
We hope this model will be useful to think about the impact of vaccine mandates
beyond this application.

A first implication of our model
is that, when we aggregate this decision rule across individuals,
the mandate's total effect $E(M)$ at the deadline can be decomposed into
two channels.
One is the pull-forward channel, $A(M)$, and the other is an
induction channel, $I(M)$, so that
\[ E(M) = A(M) + I(M) \]
A wider and more technical discussion of this point can be found in
Appendix~\ref{appx:decomp}.

%The time-limited mandate adds a one-off payoff $M$ (for example, access to a 
%memorable event) at a deadline $T_0$, and heterogeneity in an ``act-now'' 
%threshold $Z$ summarises heterogeneity in perceived costs and present bias. 

A second intuitive implication of the model is that the only people
whose vaccination can be \textit{induced}\textemdash rather than pulled
forward\textemdash by the mandate are those who do not vaccinate in
the long run, absent the mandate\footnote{
    More precisely, those who do not vaccinate in
    the long run, after the mandate date, absent the mandate.}.
Empirically, this share of people is estimated by the share of unvaccinated
people in the control group as the uptake growth rate goes to zero, i.e.
by the steady-state share of unvaccinated in the control group.
In our target population\textemdash similarly to the rest of WA\textemdash
the untreated group (Year 11 students) had an uptake above 98\% at 80 days
from the mandate deadline.
This implies that less than 2\% of the target individuals was eligible to be
induced to vaccinate\textemdash among them there could be individuals
who were not recommended vaccination due to medical reasons.
In terms of the formal model, this corresponds to a small baseline
non-vaccination mass, so the maximum headroom for induction is mechanically
limited (see Equations~\eqref{eq:nv0} and~\eqref{eq:nv-drop}).
In short, while it is relatively hard to induce new vaccinations in a population
with high long-run coverage at the baseline,\footnote{
    ``Baseline'' is to be intended as ``absent the mandate of interest''.
    This implies that ``high long-run coverage'' refers to the coverage of
    the control group, which is used to build the counterfactual outcomes
    for the treated group.
}
populations where the long-run coverage is lower offer interventions a greater
chance to induce new vaccinations. Conversely, a \textit{Leavers}-type mandate
targeting populations with lower expected long-run uptake may induce new
vaccinations.

Relatedly, the model captures how other baseline incentives shape the effect
composition and, potentially, its total size.
Vaccination brings health benefits,
$eH\lambda_t$\textemdash via a reduction in the expected harm from
COVID-19 infections\textemdash and non-health benefits, $b_t$, such as the
ability to travel.
Indeed, the WA government reopening the State's ``controlled border'' was
explicitly contingent on achieving a 90\% double-dose vaccination rate among
those aged 12+ under the State's \emph{Safe Transition Plan}
\citep{WA_Safe_Transition_2021}.
It did not, however, provide a deadline.\footnote{
    On 5 November 2021, WA's government announced the 90\% threshold and
    committed to providing a reopening date once 80\% of WA's residents aged 12
    or above were vaccinated. This occurred on 13 December 2021, past the Leavers
    mandate deadline, and the reopening date was set to 5 February 2022 based
    on WA's coverage projections \citep{WA_Health_Update_2021_12_13}.
    On 18 February 2022 the reopening date was moved to 3 March 2022 due to
    the Omicron wave \citep{WA_Gov_BorderOpening_2022_03_03}.
}
This policy creates strong baseline incentives to vaccinate. %
In our model, a border reopening increases post-deadline non-health benefits
$b_t$, which raise the post-deadline one-period increments
$\Delta_t$ and hence the best post-deadline baseline increment
$\overline{\Delta}_{\text{post}}$.
From the analyst's perspective\footnote{
    Rather than the agent's perspective, who either lives in the baseline
    (no-mandate) world, or in the world with the mandate.},
comparing this richer baseline path to the
mandate scenario with a given $M$,
a higher $\overline{\Delta}_{\text{post}}$ makes it
less likely that vaccinating at
$T_0$ with the mandate payoff yields a larger one-period gain than
vaccinating later without the mandate, and thus makes it less likely that the
induction channel exists
(see Corollary~1 and Equation~\eqref{eq:induction-condition}).

%Since inducing new vaccinations depends on the share of non-vaccinated, this 
%travel policy mostly attenuates the induction channel\textemdash consistent with
%our evidence.
%This is only an example of what can drive baseline uptake upwards: the
%non-health benefits term in our model captures any incentive to vaccinate
%unrelated to Leavers and affecting both treatment and control groups.

Moreover, Western Australia at the time was successfully pursuing a
``zero-Covid'' strategy, keeping new cases in the low single digits. In our
model, this low incidence means that the health-hazard term, $\lambda_t$, is
small around the mandate deadline, so the immediate health benefit of
vaccinating is limited. This reduces the incentive for present-biased
individuals to vaccinate early, creating a larger pool of ``vaccinate later''
types who can be brought forward by the mandate payoff, $M$, rather than a large
pool of individuals who would otherwise never vaccinate.
Taken together with the strong post-deadline incentives discussed above, this
makes it unlikely that the condition for the existence of the induction
channel (Corollary~1) holds in our
setting, so that the Leavers mandate primarily accelerates vaccinations that
would have occurred later anyway.

In sum, our model offers an explanation of why the Leavers mandate
had a strong pull-forward impact but was unable to induce new vaccinations.
The 90\% uptake required to re-open the State's borders ensured high long-run
uptake, suppressing the induction effect of the mandate.
On the other hand, the lack of a deadline to reopening the borders and the
low COVID-19 incidence might have encouraged procrastination,
strengthening the pull-forward effect.
The more general implications of our model, applying to all
time-limited vaccine mandates, are discussed in the next section.

We close this section with a note on the model's interpretation.
While, for ease of exposition, in Appendix~\ref{app:mandate}
we define the model's variables in terms of objective states,
they can also be interpreted as subjective beliefs about such states while
leaving the model unchanged.
This is particularly relevant when estimating or predicting mandate impacts
in populations where there is a strong mismatch between the beliefs and facts.
For instance, if a part of the population believes that the risk of serious
illness or death from a given pathogen is low\textemdash
contrary to the best available evidence\textemdash
they will nonetheless behave as if the risk were low.

\section{Conclusion and policy implications\label{sec:conclusion}}

This paper studies how a time-limited vaccine mandate changes vaccination
behaviour---not only whether it increases uptake, but whether it does so by
\emph{inducing} vaccinations that would not otherwise occur
or by \emph{pulling forward} vaccinations that would occur later anyway
\citep{brehm2022}.
We use Western Australia's \textit{Leavers mandate} as a working example---a
low-coercion COVID-19
policy requiring recent high-school graduates to receive a first dose to access
popular post-graduation events---and estimate it caused a sharp increase in first-dose
vaccination at the mandate deadline of 9.3 percentage points.
Dynamic regression-discontinuity estimates show that virtually the entire
effect reflects pull-forward behaviour. The mandate shifted the timing of
vaccination forward by as much as 80 days for students who were
eventually going to vaccinate, while we do not detect a statistically
significant increase in long-run coverage.
In other words, this policy had a large impact on the timing of vaccinations,
accelerating them, but did not induce new vaccinations.

This distinction is central for policy evaluation and design.
When an intervention includes a deadline---the defining feature of ``time-limited
mandates'' in a strict sense---short-run
increases in uptake need not translate into higher long-run coverage
\citep{ODonoghueRabin1999}.
Accordingly, analysts should evaluate impacts beyond the deadline and, where
possible, explicitly decompose effects into induction and pull-forward
components. Failing to do so risks misclassifying accelerated vaccinations as
net gains in coverage, and therefore mis-specifying the policy objective and
its welfare consequences.

To interpret our empirical findings and clarify how similar mandates may perform
in other environments, we develop a simple model of first-dose vaccination under
present bias and a time-limited mandate, extending \cite{ODonoghueRabin1999}.
The model formalises how the same mandate can generate different effect
compositions depending on (i) baseline incentives to vaccinate around and after
the deadline and (ii) the salience and immediacy of the mandate's conditional
benefit. In our application, the model helps rationalise a large pull-forward
effect with null induction: strong long-run incentives in WA (e.g., reopening
of state borders) likely
pushed long-run uptake close to 100\%, while the state's very low COVID-19
incidence and adolescents' low perceived risk of severe disease increased
procrastination incentives. The Leavers mandate then operated primarily as an
effective deadline that overcame procrastination rather than as a mechanism for
convincing people to vaccinate at all.

Three policy implications follow.
First, low-coercion, non-monetary, and time-limited incentives can be highly
effective tools for accelerating vaccination campaigns, which is particularly
useful when the
public-health objective is earlier immunity (e.g., in fast transmission phases
or when timing is crucial).
Second, the total effect of such a mandate is greater in presence of:
lower baseline incentives to vaccinate around the deadline;
a conditional benefit that is valuable to the
target group and tied to the deadline; and behavioural frictions such as
present bias or high perceived immediate costs of vaccinating\footnote{
    This is true as long as (and up to the point where) such factors do not
    dominate vaccination incentives.}.
Unless the target population is already prone to vaccinating, so that it will
display high long-run uptake, such factors will also strengthen the
induction effect, convincing people to vaccinate.
Third, mandate effectiveness is strongly linked to the broader epidemiological
environment and policy mix: beliefs about the infection risk, the
expected private and
social benefits from vaccinating, and the concurrent policies that shift long-run
incentives all shape not just the size of the uptake response but its
composition between pull-forward and induction.

Finally, the Leavers mandate offers a pragmatic rationale for tying
vaccination to one-off large social gatherings that can become super-spreader
events. This rationale is strongest when vaccines reduce onward transmission,
due to infection externalities, but remains relevant when they primarily reduce
severe illness and the risk of hospital admission, due to hospital congestion
externalities\textemdash particularly in intensive-care units.
More generally, our results, on the one hand, suggest
that time-limited, low-coercion mandates may offer a politically
and ethically viable instrument to accelerate uptake, and, on the other hand,
highlight why impact evaluation should aim at separating pull-forward and
induction effects.

\bibliographystyle{apa}
\bibliography{vax_mandate_rdd_biblio.bib}

\begin{thebibliography}{}

\bibitem[\protect\astroncite{{ABC News}}{2021a}]{turn1search0}
{ABC News} (2021a).
\newblock Mandatory covid-19 vaccines for wa fifo and mining workers, school
  leavers.
\newblock {\em ABC News}.
\newblock Contains Dawson's statement on wristband proof; accessed May 2025.

\bibitem[\protect\astroncite{{ABC News}}{2021b}]{turn0search0}
{ABC News} (2021b).
\newblock Mark mcgowan holds firm on setting international travel date.
\newblock {\em ABC News}.
\newblock Contains statements on Year 12 vaccination requirement; accessed May
  2025.

\bibitem[\protect\astroncite{{ABC News}}{2021c}]{McGowan_Oct1_2021}
{ABC News} (2021c).
\newblock Western australia premier signals vaccine requirement for school
  leavers is `strong possibility'.
\newblock {\em ABC News}.
\newblock Accessed May 2025.

\bibitem[\protect\astroncite{{ABC South West
  WA}}{2021}]{ABC_DunsboroughLeavers2021}
{ABC South West WA} (2021).
\newblock Dunsborough prepares for vaccinated leavers as mandate bars people
  from entry.
\newblock {\em ABC News}.
\newblock Accessed May 2025.

\bibitem[\protect\astroncite{Abrevaya and Mulligan}{2011}]{Abrevaya2011}
Abrevaya, J. and Mulligan, K. (2011).
\newblock Effectiveness of state-level vaccination mandates: Evidence from the
  varicella vaccine.
\newblock {\em Journal of Health Economics}, 30(5):966--976.

\bibitem[\protect\astroncite{Acton et~al.}{2024}]{Acton2024}
Acton, R.~K., Cao, W., Cook, E.~E., Imberman, S.~A., and Lovenheim, M.~F.
  (2024).
\newblock The effect of vaccine mandates on disease spread.
\newblock {\em Journal of Human Resources}.

\bibitem[\protect\astroncite{Aslim et~al.}{2024}]{Aslim2024}
Aslim, E. et~al. (2024).
\newblock Vaccination policy, delayed care, and health expenditures.
\newblock {\em Economic Journal}.
\newblock Advance online.

\bibitem[\protect\astroncite{{Australian Government Department of
  Health}}{2021}]{DH_AustraliaRollout20211021}
{Australian Government Department of Health} (2021).
\newblock Covid-19 vaccine roll-out: 21 october 2021.
\newblock
  \url{https://www.health.gov.au/sites/default/files/documents/2021/10/covid-19-vaccine-rollout-update-21-october-2021.pdf}.
\newblock Accessed May 2025.

\bibitem[\protect\astroncite{Barber and West}{2022}]{BarberWest2022}
Barber, A. and West, J. (2022).
\newblock Conditional cash lotteries increase covid-19 vaccination rates.
\newblock {\em Journal of Health Economics}, 84:102641.

\bibitem[\protect\astroncite{Bardosh et~al.}{2022}]{bardosh2022}
Bardosh, K., de~Figueiredo, A., Gur-Arie, R., Jamrozik, E., Doidge, J.,
  Lemmens, T., et~al. (2022).
\newblock The unintended consequences of covid-19 vaccine policy: why mandates,
  passports and restrictions may cause more harm than good.
\newblock {\em BMJ Global Health}, 7:e008684.

\bibitem[\protect\astroncite{Brehm et~al.}{2022}]{brehm2022}
Brehm, M.~E., Brehm, P.~A., and Saavedra, M. (2022).
\newblock The ohio vaccine lottery and starting vaccination rates.
\newblock {\em American Journal of Health Economics}, 8(3):387--411.

\bibitem[\protect\astroncite{Buttenheim et~al.}{2022}]{Buttenheim2022}
Buttenheim, A., Milkman, K.~L., Duckworth, A.~L., Gromet, D.~M., Patel, M., and
  Chapman, G. (2022).
\newblock Effects of ownership text message wording and reminders on receipt of
  an influenza vaccination: A randomized clinical trial.
\newblock {\em JAMA Network Open}, 5(2):e2143388--e2143388.

\bibitem[\protect\astroncite{Calonico et~al.}{2024}]{rdrobustManual}
Calonico, S., Cattaneo, M.~D., Farrell, M.~H., and Titiunik, R. (2024).
\newblock {\em rdrobust: Software for Regression-Discontinuity Designs}.
\newblock rdpackages project.
\newblock Section “Options”, page 5.

\bibitem[\protect\astroncite{Calonico et~al.}{2014}]{Calonico2014}
Calonico, S., Cattaneo, M.~D., and Titiunik, R. (2014).
\newblock Robust nonparametric confidence intervals for
  regression-discontinuity designs.
\newblock {\em Econometrica}, 82(6):2295--2326.

\bibitem[\protect\astroncite{Campos-Mercade et~al.}{2024}]{Campos-Mercade2024}
Campos-Mercade, P., Meier, A.~N., Meier, S., Pope, D.~G., Schneider, F.~H., and
  Wengström, E. (2024).
\newblock Incentives to vaccinate.
\newblock Working Paper 32899, National Bureau of Economic Research.

\bibitem[\protect\astroncite{Carlson et~al.}{2023}]{Carlson2023}
Carlson, S.~J., Attwell, K., Roberts, L., Hughes, C., and Blyth, C.~C. (2023).
\newblock West australian parents’ views on vaccinating their children
  against covid-19: a qualitative study.
\newblock {\em BMC Public Health}, 23(1):1764.

\bibitem[\protect\astroncite{Carlson et~al.}{2022}]{Carlson2022}
Carlson, S.~J., McKenzie, L., Roberts, L., Blyth, C.~C., and Attwell, K.
  (2022).
\newblock Does a major change to a covid-19 vaccine program alter vaccine
  intention? a qualitative investigation.
\newblock {\em Vaccine}, 40(4):594--600.

\bibitem[\protect\astroncite{Carmody}{2021}]{Carmody_Oct5_2021}
Carmody, J. (2021).
\newblock Mandatory covid-19 vaccines for wa fifo and mining workers, school
  leavers.
\newblock {\em ABC News}.
\newblock Accessed May 2025.

\bibitem[\protect\astroncite{Carmody}{2019}]{ABC2019}
Carmody, R. (2019).
\newblock Parents fighting to hold back their child from starting school in wa
  forced to consider drastic action.
\newblock {\em {ABC News}}.
\newblock Accessed: 14 May 2025.

\bibitem[\protect\astroncite{Carpenter and Lawler}{2019}]{Carpenter2019}
Carpenter, C.~S. and Lawler, E.~C. (2019).
\newblock Direct and spillover effects of middle school vaccination
  requirements.
\newblock {\em American Economic Journal: Economic Policy}, 11(1):95--125.

\bibitem[\protect\astroncite{Cattaneo et~al.}{2015}]{Cattaneo2015}
Cattaneo, M.~D., Frandsen, B.~R., and Titiunik, R. (2015).
\newblock Randomization inference in the regression discontinuity design: An
  application to party advantages in the {U}{S} senate.
\newblock {\em Journal of Causal Inference}, 3(1):1--24.

\bibitem[\protect\astroncite{Cattaneo and Titiunik}{2022}]{cattaneoReview}
Cattaneo, M.~D. and Titiunik, R. (2022).
\newblock Regression discontinuity designs.
\newblock {\em Annual Review of Economics}, 14(Volume 14, 2022):821--851.

\bibitem[\protect\astroncite{Chang}{2016}]{Chang2016}
Chang, L.~V. (2016).
\newblock The effect of state insurance mandates on infant immunization rates.
\newblock {\em Health Economics}, 25(3):372--386.

\bibitem[\protect\astroncite{Fitzpatrick et~al.}{2023}]{Fitzpatrick2023}
Fitzpatrick, H. et~al. (2023).
\newblock The impact of provincial proof-of-vaccination policies on
  age-specific first-dose uptake.
\newblock {\em Health Affairs}, 42(3):e202201237.

\bibitem[\protect\astroncite{Freedman et~al.}{2022}]{Freedman2023}
Freedman, S.~M., Sacks, D.~W., Simon, K.~I., and Wing, C. (2022).
\newblock Direct and indirect effects of vaccines: Evidence from covid-19.
\newblock Working Paper 30550, National Bureau of Economic Research.

\bibitem[\protect\astroncite{Gebremariam et~al.}{2025}]{Gebremariam2025}
Gebremariam, A.~G., Genie, M., Le, H., Attwell, K., Liu, B., Regan, A.~K.,
  Beard, F.~H., Macartney, K., Paolucci, F., Moore, H.~C., and Blyth, C.~C.
  (2025).
\newblock Impact of vaccine mandates and removals on covid-19 vaccine uptake in
  australia and international comparators: a study protocol.
\newblock {\em BMJ Open}, 15(7).

\bibitem[\protect\astroncite{{Government of Western
  Australia}}{2021}]{WA_Safe_Transition_2021}
{Government of Western Australia} (2021).
\newblock Wa's safe transition plan.
\newblock
  \url{https://www.wa.gov.au/government/announcements/was-safe-transition-plan}.
\newblock Published 5 Nov 2021; last updated 14 Mar 2022. Accessed 20 Oct 2025.

\bibitem[\protect\astroncite{{Government of Western
  Australia}}{2022}]{WA_Gov_BorderOpening_2022_03_03}
{Government of Western Australia} (2022).
\newblock Wa's border opening from thursday 3 march 2022.
\newblock
  \url{https://www.wa.gov.au/government/announcements/was-border-opening-thursday-3-march-2022}.

\bibitem[\protect\astroncite{Hahn et~al.}{2001}]{hahn2001}
Hahn, J., Todd, P., and der Klaauw, W.~V. (2001).
\newblock Identification and estimation of treatment effects with a
  regression-discontinuity design.
\newblock {\em Econometrica}, 69(1):201--209.

\bibitem[\protect\astroncite{Lawler}{2017}]{Lawler2017}
Lawler, E.~C. (2017).
\newblock Effectiveness of vaccination recommendations versus mandates:
  Evidence from the hepatitis a vaccine.
\newblock {\em Journal of Health Economics}, 52:45--62.

\bibitem[\protect\astroncite{McCrary}{2008}]{mccrary08}
McCrary, J. (2008).
\newblock Manipulation of the running variable in the regression discontinuity
  design: A density test.
\newblock {\em Journal of Econometrics}, 142(2):698--714.
\newblock The regression discontinuity design: Theory and applications.

\bibitem[\protect\astroncite{Nguyen et~al.}{2024}]{Nguyen2024}
Nguyen, M.-H., Hoang, V.-N., Nghiem, S., and Nguyen, L.~A. (2024).
\newblock The dynamic and heterogeneous effects of covid-19 vaccination
  mandates in the usa.
\newblock {\em Health Economics}.
\newblock Forthcoming.

\bibitem[\protect\astroncite{O'Donoghue and Rabin}{1999}]{ODonoghueRabin1999}
O'Donoghue, T. and Rabin, M. (1999).
\newblock Doing it now or later.
\newblock {\em American Economic Review}, 89(1):103--124.

\bibitem[\protect\astroncite{{Office of the Auditor General,
  WA}}{2021}]{WA_VaccinationProgramAudit2021}
{Office of the Auditor General, WA} (2021).
\newblock Wa's covid-19 vaccine roll-out.
\newblock Technical report, Government of Western Australia.
\newblock Accessed May 2025.

\bibitem[\protect\astroncite{{Parliament of New South
  Wales}}{}]{nsw_education_act_1990}
{Parliament of New South Wales}.
\newblock Education act 1990 (nsw).
\newblock
  \url{https://legislation.nsw.gov.au/view/whole/html/inforce/current/act-1990-008}.
\newblock Compulsory school-age provisions (section 21B); Accessed 12 June
  2025.

\bibitem[\protect\astroncite{{School Curriculum and Standards
  Authority}}{2021}]{scsa2021}
{School Curriculum and Standards Authority} (2021).
\newblock {11to12 Circulars - Edition 4, May 2021}.
\newblock (Accessed 22 May 2025).

\bibitem[\protect\astroncite{{Schoolies.com}}{2025}]{schoolies2025}
{Schoolies.com} (2025).
\newblock Leavers: The biggest graduation celebration for year 12s in the
  dunsborough area.
\newblock Accessed: 11 June 2025.

\bibitem[\protect\astroncite{Staff and
  Agencies}{2021}]{Guardian_ReopeningRoadmap2021}
Staff and Agencies (2021).
\newblock Australia to pass 80\% vaccination target today, pm says; wa
  reopening roadmap revealed.
\newblock {\em \em The Guardian}.
\newblock Accessed May 2025.

\bibitem[\protect\astroncite{{Tribe
  Travel}}{2021}]{TribeTravel_LeaversInfo2021}
{Tribe Travel} (2021).
\newblock Faq: About leavers wa.
\newblock \url{https://www.tribetravel.com.au/about-leavers-wa}.
\newblock Accessed May 2025.

\bibitem[\protect\astroncite{{WA Department of
  Health}}{}]{HealthyWA_Case_6Oct2021}
{WA Department of Health}.
\newblock Fact sheet: Covid-19 case numbers as of 6 october 2021.
\newblock
  \url{https://www.healthywa.wa.gov.au/~/media/Corp/Documents/Health-for/Infectious-disease/COVID19/COVID19-Aboriginal-Sector-Communications-Update-28.pdf}.
\newblock Accessed May 2025.

\bibitem[\protect\astroncite{{WA Department of Health,
  PHEOC}}{2021}]{WA_LeaversProof2021}
{WA Department of Health, PHEOC} (2021).
\newblock School leavers required to show proof of covid-19 vaccination.
\newblock
  \url{https://www.wa.gov.au/government/announcements/school-leavers-required-show-proof-of-covid-19-vaccination}.
\newblock Accessed May 2025.

\bibitem[\protect\astroncite{{WA
  Government}}{2021a}]{WA_VaccineProgramPhases2021}
{WA Government} (2021a).
\newblock First for covid-19 vaccine and vaccination hubs in wa.
\newblock
  \url{https://www.health.wa.gov.au/Media-releases/2021/First-for-COVID-19-vaccine-and-vaccination-hubs-in-WA}.
\newblock Accessed May 2025.

\bibitem[\protect\astroncite{{WA
  Government}}{2021b}]{WestAust_VaxMandateAnnounce2021}
{WA Government} (2021b).
\newblock Mandatory vaccinations for dunsborough school leavers event.
\newblock
  \url{https://www.wa.gov.au/government/announcements/mandatory-vaccinations-dunsborough-school-leavers-event}.
\newblock Published Nov 2021; accessed May 2025.

\bibitem[\protect\astroncite{{Western Australia Department of
  Health}}{2021a}]{wahealth2021nov01}
{Western Australia Department of Health} (2021a).
\newblock {COVID-19 update 1 November 2021}.
\newblock New COVID-19 cases reported in WA: 0. Government of Western
  Australia. Accessed 22 May 2025.

\bibitem[\protect\astroncite{{Western Australia Department of
  Health}}{2021b}]{WA_Health_Update_2021_12_13}
{Western Australia Department of Health} (2021b).
\newblock Covid-19 update 13 december 2021.
\newblock
  \url{https://www.health.wa.gov.au/Media-releases/2021/COVID-19-update-13-December-2021}.

\bibitem[\protect\astroncite{{Western Australia Department of
  Health}}{2021c}]{wahealth2021sep25}
{Western Australia Department of Health} (2021c).
\newblock {COVID-19 update 25 September 2021}.
\newblock New COVID-19 cases reported in WA: 0. Government of Western
  Australia. Accessed 22 May 2025.

\bibitem[\protect\astroncite{{Western Australian Department of
  Education}}{}]{WAEnrollingSchool}
{Western Australian Department of Education}.
\newblock Enrol at a western australian school.
\newblock \url{https://www.education.wa.edu.au/enrolling-in-school}.
\newblock Accessed: 2025-05-14.

\bibitem[\protect\astroncite{{Western Australian Department of
  Education}}{2022}]{WAEnrolmentPolicy}
{Western Australian Department of Education} (2022).
\newblock Enrolment in public schools policy.
\newblock \url{https://www.education.wa.edu.au/dl/4mn0ozv}.
\newblock Effective 18 July 2022; Version 3.0; accessed 2025-05-14.

\end{thebibliography}

\clearpage
\appendix

\section{Extra tables}

\begin{table}[!h]
  \centering
  \caption{\label{tab:rdd_cont}Sharp RDD Estimates (continuity assumption)}
  \centering
  \begin{threeparttable}
    \begin{tabular}[t]{lcccc}
      \toprule
      \multicolumn{1}{c}{ }       & \multicolumn{2}{c}{12-month bandwidth} & \multicolumn{2}{c}{Auto bandwidth}                                        \\
      \cmidrule(l{3pt}r{3pt}){2-3} \cmidrule(l{3pt}r{3pt}){4-5}
                                  & First mention                          & Deadline                           & First mention     & Deadline         \\
      \midrule
      \textit{Attending students} &                                        &                                    &                   &                  \\
      \hspace{3mm}Coeff.          & 0.000                                  & 0.093                              & 0.028             & 0.074            \\
      \hspace{3mm}Std. error      & (0.009)                                & (0.008)                            & (0.026)           & (0.023)          \\
      \hspace{3mm}95\% CI         & {}[-0.018, 0.017]                      & {}[0.078, 0.108]                   & {}[-0.023, 0.079] & {}[0.028, 0.120] \\
      \hspace{3mm} $N_{-}, N_{+}$ & 24313, 22108                           & 24313, 22108                       & 3957, 5690        & 3957, 5690       \\
      \textit{General population} &                                        &                                    &                   &                  \\
      \hspace{3mm}Coeff.          & -0.002                                 & 0.082                              & 0.027             & 0.067            \\
      \hspace{3mm}Std. error      & (0.008)                                & (0.007)                            & (0.022)           & (0.022)          \\
      \hspace{3mm}95\% CI         & {}[-0.017, 0.013]                      & {}[0.068, 0.096]                   & {}[-0.017, 0.071] & {}[0.024, 0.110] \\
      \hspace{3mm} $N_{-}, N_{+}$ & 30391, 30215                           & 30391, 30215                       & 5024, 7540        & 5024, 7540       \\
      \bottomrule
    \end{tabular}
    \begin{tablenotes}
      \item \textit{Notes}: This table shows the RDD coefficients estimating
      the effect of the Leavers' mandate on vaccination rates. The cutoff we use
      to divide the sample between treatment and control group is date of birth
      30 June 2004. In columns 1 and 2, we use our preferred bandwidth, 12 months.
      This implies comparing the full cohorts of Year-12 and Year-11 students (the
      \textit{Attending students} results), or comparing people of Year-12 age
      with people of Year-11 age  (the \textit{General population} results).
      Moreover, columns 1 and 3 report the (placebo) impacts on the day in which
      the mandate was first mentioned publicly, and columns 2 and 4 report the impact
      of the policy at its deadline.
      In columns 3 and 4, we use the ``automatic'' bandwidth, by which we mean the
      bandwidth that minimises the asymptotic mean-squared error of the
      local-polynomial RD point estimate, as derived in \cite{Calonico2014}.
      This bandwidth is equal to 2, so that only people born $\pm 2$ months from the
      cutoff are included in the estimation sample.
      We estimate standard errors via a heteroskedasticity-robust nearest neighbor
      variance estimator that uses a minimum of 3 neighbours
      \citep{rdrobustManual}.
    \end{tablenotes}
  \end{threeparttable}
\end{table}

\begin{table}[!h]
  \centering
  \caption{\label{tab:rdd_randinf}Sharp RDD Estimates (local randomisation)}
  \centering
  \begin{threeparttable}
    \begin{tabular}[t]{lccc}
      \toprule
                                      & Diff. in means & Exact p-value & Asympt. p-value \\
      \midrule
      1-month window - Deadline       & 0.090          & 0.000         & 0.000           \\
      1-month window - First mention  & 0.020          & 0.154         & 0.138           \\
      6-month window - Deadline       & 0.104          & 0.000         & 0.000           \\
      6-month window - First mention  & 0.006          & 0.331         & 0.323           \\
      12-month window - Deadline      & 0.112          & 0.000         & 0.000           \\
      12-month window - First mention & 0.015          & 0.001         & 0.001           \\
      \bottomrule
    \end{tabular}
    \begin{tablenotes}
      \item \textit{Notes}: This table shows the local-randomisation RDD
      coefficients \citep{Cattaneo2015} estimating
      the effect of the Leavers' mandate on vaccination rates.
      The cutoff we use
      to divide the sample between treatment and control group is date of birth
      30 June 2004.
      In Column 1, we report the coefficient (difference in means), in Column 2
      the exact p-value, and in Column 3 the asymptotic p-value.
      The window sizes are 1 in rows 1-2, 6 in rows 3-4, and 12 in rows 5-6.
      Even-numbered rows report impacts at the date of the first mention of the mandate, while
      odd-numbered rows report impact at deadlines.

    \end{tablenotes}
  \end{threeparttable}
\end{table}

\newpage

\section{Toy model of vaccination decisions}\label{app:mandate}

We adapt \citet{ODonoghueRabin1999}'s procrastination model to vaccination with
a finite horizon of $T$ periods and a single action, taking the first COVID-19
dose, with immediate cost $c\!\ge\!0$ and delayed benefits. Individuals have
quasi-hyperbolic preferences with present-bias $\beta\in(0,1]$ and $\delta=1$
(all results extend to $\delta<1$).

\subsection{Primitives}
Time is discrete, $t\in\{0,1,\dots,T\}$, and a single one-off action
(vaccination) is available in each period until it is taken.
Taking the action at $t$ has an immediate cost $c\ge 0$.
Let $\lambda_s$ denote the period-$s$ health hazard\textemdash
severe illness or death\textemdash
faced by the unvaccinated,
with the vaccine reducing this hazard by a constant factor $e\in[0,1]$, and
let $H>0$ denote the harm associated with each adverse event.
Let $b_s$ capture non-mandate, non-health per-period benefits
that accrue to being vaccinated by calendar time $s$
(e.g., work/travel convenience, peace of mind).
A policy mandate specifies a deadline $T_0\le T$ and provides a one-off
payoff $M>0$ if vaccination occurs by $T_0$.

\subsection{Continuation value and one-period increment}
The continuation benefit of vaccinating at time $\tau$ is
\[
    y_\tau \;\equiv\; \sum_{s=\tau}^{T} \big( eH\,\lambda_s + b_s \big),
\]
so the \emph{baseline one-period incremental benefit} is
\[
    \Delta_t \;\equiv\; y_t - y_{t+1} \;=\; eH\,\lambda_t + b_t,
\]
where $eH\,\lambda_t$ is the \emph{expected health harm avoided in period $t$}
by vaccinating at $t$ rather than $t{+}1$, i.e.
vaccine efficacy $e$ times harm-per-event $H$ times baseline risk $\lambda_t$.

With the mandate, the continuation benefit is $y_\tau^{M} = y_\tau + M\,\mathbbm{1}\{\tau\le T_0\}$ and the one-period increment becomes
\[
    \Delta_t^{M} \;=\; \Delta_t + M\,\mathbbm{1}\{t=T_0\}.
\]

\subsection{Decision rule}
At time $t$, present-biased individuals compare \emph{vaccinate now} versus \emph{wait one period}:
\[
    U_t^{\text{now}} = -c + \beta\,y_t,\qquad U_t^{\text{wait}} = -\beta c + \beta\,y_{t+1}.
\]
Define the \emph{action threshold}
\begin{equation}
    Z \;\equiv\; \frac{1-\beta}{\beta}\,c.
    \label{eq:threshold}
\end{equation}
\textbf{Lemma (Decision rule).} A person vaccinates at $t$ iff
$\Delta_t^{M} \ge Z$. In particular, at the deadline $t=T_0$ the rule is
\begin{equation}
    \Delta_{T_0} + M \;\ge\;  Z .
    \label{eq:deadline-cut}
\end{equation}

Define also the best pre-deadline baseline increment
\[
    \overline{\Delta}_{\text{pre}} \;\equiv\; \sup_{0\le s \le T_0} \Delta_s .
\]

\noindent\textbf{Assumption 1 (Monotone pre-deadline increments).}
The baseline one-period increments are weakly increasing up to the deadline:
\[
    \Delta_s \;\le\; \Delta_{T_0}\quad\text{for all } s \le T_0,
\]
so that $\overline{\Delta}_{\text{pre}} = \Delta_{T_0}$.
Intuitively, the one-period advantage from being vaccinated does not fall as
the mandate deadline approaches.
In our application, this assumption is realistic: prior to the Leavers
deadline, infection risk and non-health benefits of vaccination (such as
general travel convenience) did not decrease over time, while the major
policy change affecting $b_t$ (the border reopening tied to 90\% coverage)
occurred only after $T_0$.

\subsection{Total effect and channel decomposition \label{appx:decomp}}

\noindent\textbf{Proposition 1 (Decomposition).}
Decompose the total effect $E(M)$ as $E(M)=A(M)+I(M)$, where
\begin{align}
    A(M)\; & =\;
    \max\Big\{0,\;
    F_Z\!\Big(\min\{\Delta_{T_0}+M,\ \overline{\Delta}_{\text{post}}\}\Big)
    - F_Z\!\big(\Delta_{T_0}\big)
    \Big\}
           &                 & \text{\emph{(pull-forward)}} ,
    \label{eq:accel}                                          \\
    I(M)\; & =\; E(M) - A(M)
           &                 & \text{\emph{(induction)}}.
    \label{eq:induct}
\end{align}

In words, the mandate raises the cutoff value of $Z$ for vaccinating at the
deadline from $\Delta_{T_0}$ to $\Delta_{T_0}+M$.
Individuals are indexed by their threshold $Z$. Among those
with $\Delta_{T_0} < Z \le \Delta_{T_0}+M$, the mandate makes vaccination at
$T_0$ worthwhile. Those with
$Z\in\big(\Delta_{T_0},\,\min\{\Delta_{T_0}+M,\overline{\Delta}_{\text{post}}\}\big]$
would, absent the mandate, eventually vaccinate at some post-deadline time
(because $Z\le\overline{\Delta}_{\text{post}}$), so the mandate pulls
their vaccination forward to $T_0$. This gives the pull-forward mass $A(M)$ in
\eqref{eq:accel}. The
$\min\{\Delta_{T_0}+M,\overline{\Delta}_{\text{post}}\}$ term reflects that only
individuals who would have vaccinated at some post-deadline time in the
baseline ($Z\le\overline{\Delta}_{\text{post}}$) can be classified as
pull-forward rather than induced. When $\overline{\Delta}_{\text{post}}
    \le\Delta_{T_0}$ this set is empty, and $\max\{0,\cdot\}$ in
\eqref{eq:accel} sets $A(M)=0$, so no vaccination is accelerated.

Those with thresholds above this range but still below the deadline cutoff, \newline
$Z\in\big(\max\{\Delta_{T_0},\overline{\Delta}_{\text{post}}\},\,
    \Delta_{T_0}+M\big]$,
would not reach their threshold at or after $T_0$ in the baseline\textemdash
and hence would not vaccinate at or after the deadline\textemdash
but the deadline payoff now makes vaccination worthwhile.
These individuals are \emph{induced} by the mandate,
and their mass is captured by $I(M)=E(M)-A(M)$. Corollary~1 shows that
induction arises iff $\Delta_{T_0}+M>\overline{\Delta}_{\text{post}}$, i.e.
whenever vaccinating at the deadline with the mandate payoff is more attractive
than vaccinating at the best post-deadline time in the baseline scenario.

\noindent \textbf{Corollary 1 (Existence of induction).} Induction arises iff
\begin{equation}
    \Delta_{T_0} + M \;>\; \overline{\Delta}_{\text{post}} .
    \label{eq:induction-condition}
\end{equation}
Otherwise the mandate only accelerates: $E(M)=A(M)$.

\noindent \textbf{Remark 1 (positive total effect).}
From Proposition 1, $E(M)>0$ iff $F_Z(\Delta_{T_0}+M)>F_Z(\Delta_{T_0})$,
i.e. iff $\Pr\!\big(Z\in(\Delta_{T_0},\,\Delta_{T_0}+M]\big)>0$.
If $F_Z$ admits a density $f_Z$ that is positive at $\Delta_{T_0}$,
then $E(M)\approx f_Z(\Delta_{T_0})\,M>0$ for all small $M>0$.
If there is a support gap $(\Delta_{T_0},\,\Delta_{T_0}+\varepsilon]$,
then $E(M)=0$ for $M\le\varepsilon$.

\subsection{Non-vaccination}
Define the best baseline one-period increment attainable at any time as
\[
    \overline{\Delta} \;\equiv\; \sup_{0\le t\le T} \Delta_t.
\]
An individual never vaccinates over the horizon iff their threshold exceeds the
best attainable increment even after the mandate payoff, $M$, namely
\[
    Z\;>\;\max\{\overline{\Delta},\,\Delta_{T_0}+M\}.
\]
Consequently, the mass of non-vaccination with mandate size $M$ is
\begin{equation}
    \mathrm{NV}(M) \;=\; 1 - F_Z\!\big(\max\{\overline{\Delta},\,\Delta_{T_0}+M\}\big),
    \label{eq:nvM}
\end{equation}
while without the mandate
\begin{equation}
    \mathrm{NV}(0) \;=\; 1 - F_Z(\overline{\Delta}).
    \label{eq:nv0}
\end{equation}
The reduction in non-vaccination due to the mandate is therefore
\begin{equation}
    \mathrm{NV}(0)-\mathrm{NV}(M) \;=\; \max\big\{0,\; F_Z(\Delta_{T_0}+M) - F_Z(\overline{\Delta})\big\}.
    \label{eq:nv-drop}
\end{equation}

Non-vaccination arises for individuals with high $Z$\textemdash that is, with a
large immediate cost $c$ and/or strong present bias (small $\beta$)\textemdash
relative to the largest one-period increment attainable at any time,
$\overline{\Delta}$. Raising the health risk $\lambda_t$
(or non-health benefits $b_t$) increases $\overline{\Delta}$ and reduces
non-vaccination even without mandates, while raising $M$ reduces it only
when

\begin{equation}
    \Delta_{T_0}+M>\overline{\Delta}
    \label{eq:nv_comp_stats}
\end{equation}

that is, when the one-period gain from vaccinating at the deadline with the
mandate payoff exceeds the best baseline one-period gain attainable at any
time. Note that $\overline{\Delta} = \max\{\overline{\Delta}_{\text{pre}},
    \overline{\Delta}_{\text{post}}\}$ and, under Assumption~1,
$\overline{\Delta}_{\text{pre}}=\Delta_{T_0}$ so that
$\overline{\Delta} = \max\{\Delta_{T_0},\overline{\Delta}_{\text{post}}\}$.
Under Assumption~1, the next lemma shows that the reduction in non-vaccination
$\mathrm{NV}(0)-\mathrm{NV}(M)$ coincides with the induction component $I(M)$ in
Proposition~1, so condition~\eqref{eq:nv_comp_stats} characterises
when the mandate generates a strictly positive long-run net increase in
ever-vaccinated individuals.

\noindent\textbf{Lemma 2 (Induction and long-run non-vaccination).}
Under Assumption~1, for every $M\ge 0$ the induction component $I(M)$ in
Proposition~1 coincides with the reduction in long-run non-vaccination:
\[
    I(M) \;=\; \mathrm{NV}(0) - \mathrm{NV}(M).
\]

\emph{Sketch of proof.}
Combining \eqref{eq:accel}--\eqref{eq:induct} with
\eqref{eq:nv-drop} and using
$\overline{\Delta} = \max\{\Delta_{T_0},\overline{\Delta}_{\text{post}}\}$,
one can check case-by-case that the set of induced individuals is
$\{Z : \overline{\Delta} < Z \le \Delta_{T_0}+M\}$, whose mass is
$F_Z(\Delta_{T_0}+M) - F_Z(\overline{\Delta})$.
By \eqref{eq:nv-drop} this equals $\mathrm{NV}(0)-\mathrm{NV}(M)$.
\hfill$\square$

\iffalse
    \subsection{Beliefs and perceived quantities (pure reinterpretation)}
    All primitives are to be read as the individual's \emph{beliefs} at the time of choice, not objective states. Specifically, $\lambda_s$ is the perceived unvaccinated hazard at calendar time $s$, $e$ is the believed relative risk reduction from vaccination, $H$ is the perceived harm scale, and $b_s$ collects perceived non-health benefits of being vaccinated by $s$. Likewise, $c$ is the perceived immediate cost of vaccinating now and $M$ is the perceived value of meeting the deadline $T_0$. Under this interpretation, $y_\tau$, $\Delta_t$, and $\Delta_t^{M}$ are \emph{expected} continuation benefits and one-period advantages computed under those beliefs. No equations change: the decision rule compares the perceived immediate advantage $\Delta_t^{M}$ to the act-now threshold $Z$, and the population response $E(M)$ and its decomposition $A(M),I(M)$ describe behavior \emph{given beliefs}. Information treatments or belief revisions correspond to replacing $(\lambda_s,e,b_s)$ with updated perceived paths; all comparative statics carry through verbatim.
\fi

\subsection{Comparative statics}
First, consider how varying the deadline payoff $M$ affects the total effect.
From Remark 1, in absence of support gaps and even for small $M>0$, the mandate
has a positive but modest impact and the composition is almost entirely
\emph{acceleration}, since the induction condition \eqref{eq:induction-condition}
does not yet bind. As $M$ increases further, the response continues to rise,
and once $\Delta_{T_0}+M$ crosses $\overline{\Delta}_{\text{post}}$ the
additional mass captured is \emph{induction}.

Other primitives shape the total effect through the baseline increments
$\Delta_t=eH\,\lambda_t+b_t$. Raising the health hazard $\lambda_t$
(e.g., moving from LR to HR periods) or increasing the non-health benefits $b_t$
shifts $\Delta_t$ upward; a higher vaccine efficacy $e$ amplifies the health
component $eH\lambda_t$. These changes typically lift both the deadline increment
$\Delta_{T_0}$ and the post-deadline peak $\overline{\Delta}_{\text{post}}$.
Holding $M$ fixed, stronger baseline incentives push more individuals above
their thresholds even without the mandate, leaving fewer at the margin at
$T_0$: the \emph{total} mandate response $E(M)$ tends to be smaller, and the
\emph{induction share} $I(M)/E(M)$ tends to shrink, so the policy acts mainly
by pulling forward the vaccinations of the remaining procrastinators.
The precise magnitude depends on where the increases in $\lambda_t$ or
$b_t$ occur in time: increases concentrated well after the deadline raise
$\overline{\Delta}_{\text{post}}$ more than $\Delta_{T_0}$, while increases
concentrated before the deadline primarily reduce the pool arriving at $T_0$
unvaccinated. In all cases, the key factor affecting the composition of the total
effect is whether $\Delta_{T_0}+M$ exceeds $\overline{\Delta}_{\text{post}}$.

\subsection{External validity \label{app:ev}}

In our data, the untreated (Year--11) control group reaches coverage
above $98\%$ by day $T_0{+}80$. In the model, this corresponds to
$F_Z(\overline{\Delta})\approx 0.98$, so the baseline non-vaccination mass
is $\mathrm{NV}(0)=1-F_Z(\overline{\Delta})\approx 0.02$ (see \ref{eq:nv0}).
Hence, the \emph{maximum} headroom for induction is about two percentage points,
since $I(M)\le \mathrm{NV}(0)$ with equality only when
$\Delta_{T_0}+M>\overline{\Delta}$ (see Equations \ref{eq:induct}
and~\ref{eq:nv-drop}). After removing individuals who are medically
ineligible (not policy-eligible), the feasible induction mass is smaller still.
Moreover, if $\Delta_{T_0}+M\le \overline{\Delta}_{\text{post}}$,
then $I(M)=0$ and any mandate effect is \emph{entirely} acceleration,
not induction (see Equations~\ref{eq:accel} and~\ref{eq:induct}). By contrast,
in populations with lower expected uptake (smaller $F_Z(\overline{\Delta})$),
the potential induction mass $1-F_Z(\overline{\Delta})$ is larger, and a
\emph{Leavers}-type mandate can lead to greater induction effects once
$\Delta_{T_0}+M>\overline{\Delta}_{\text{post}}$.
Consistent with this rationale,
we do not detect induction in our setting.

Making sense more broadly of our estimated 9.3 p.p. fully pull-forward mandate
impact\textemdash in our model, the Leavers deadline payoff \(M\)
primarily pulls forward vaccinations (rather than inducing them) when
\(\Delta_{T_0}+M\le \overline{\Delta}_{\text{post}}\)
(see Equations~\ref{eq:accel} and~\ref{eq:induct}). Western Australia's
coverage-contingent reopening (a 90\% double-dose target for ages
12+) created a travel-linked non-health benefit captured by \(b_t\),
with the border opening on 3 March 2022.
Because this benefit arrived after the Leavers deadline,
it raises \(\overline{\Delta}_{\text{post}}\) relative to \(\Delta_{T_0}\),
shrinking the mandate margin and attenuating induction. At the same time,
low COVID-19 incidence around the deadline kept the health component
\(eH\lambda_{T_0}\) small, encouraging procrastination until the deadline
payoff. Together, these features rationalize our estimate of a \(\approx 10\)
percentage point mandate effect that is fully pull-forward (acceleration only).

\end{document}